\documentstyle[12pt,epsfig]{article}

 1
\def\GeV{{\rm\, GeV}}

\def\beq{\begin{equation}}
\def\eeq{\end{equation}}
\def\bea{\begin{eqnarray}}
\def\eea{\end{eqnarray}}

\def\({\left(}   
\def\){\right)}   

\def\eq#1{{Eq.~(\ref{#1})}}

\def\npb#1#2#3{    {\it Nucl. Phys. }{\bf B#1} (19#2) #3}
\def\plb#1#2#3{    {\it Phys. Lett. }{\bf B#1} (19#2) #3}
\def\prd#1#2#3{    {\it Phys. Rev. }{\bf D#1} (19#2) #3}

\def\zpc#1#2#3{    {\it Z. Phys. }{\bf C#1} (19#2) #3}

\def\epj#1#2#3{    {\it Eur. Phys. J. }{\bf C#1} (19#2) #3}
\def\ijmpa#1#2#3{  {\it Int. J. of Mod. Phys.}{\bf A#1} (19#2) #3}

\begin{document}
%

\begin{titlepage}

\begin{center}
{\LARGE\bf{AN INVESTIGATION OF}}\\[2ex]
{\LARGE\bf{THE HARD CONTRIBUTION}}\\[2ex]
{\LARGE\bf{TO $\phi$ PHOTOPRODUCTION}}\\[6ex]

{\large \bf { Erasmo Ferreira${}^{(1)*}$\footnotetext{ ${}^*$ E-mail:
erasmo@if.ufrj.br}
  and 
 Uri Maor${}^{(1)(2)\star}$
\footnotetext{${}^{\star}$ E-mail: maor@post.tau.ac.il}}} \\[2.5ex]
{${}^{(1)}$\it Instituto de Fisica, Universidade Federal do Rio de
Janeiro}\\
{\it Rio de Janeiro RJ21945-970, BRASIL}\\[1.5ex] 
{${}^{(2)}$\it School of Physics and Astronomy, Tel Aviv University}\\
{\it Ramat Aviv 69978, ISRAEL}\\[10ex]
\end{center}
{\large \bf Abstract:}
We investigate the possibility 
that the process of $\phi$ photoproduction may have  a significant hard 
perturbative QCD component. This suggestion is based on a study 
of the energy dependence of the forward $\phi$ photoproduction cross
section followed by a calculation where we show that a coherent sum of the
pQCD and conventional soft Pomeron contributions provides an
excellent reproduction of the experimental data. Our results suggest that
the transition from the predominantly soft photoproduction of light 
$\rho$ and $\omega$ vector mesons to the predominantly hard
photoproduction of heavy $J/\Psi$ and $\Upsilon$ is smooth and gradual, 
similar to the transition observed in deep inelastic scattering studies 
of the proton structure function in the small x limit. Our predictions 
for higher HERA energies are presented.

\vspace{1cm}


\end{titlepage}

\section{Introduction}

Over the past few years we have seen a vigorous phenomenological
investigation of the Pomeron through the study of hadronic total, elastic
and diffractive cross sections, as well as the study of the proton 
deep inelastic scattering  
(DIS) structure function. In particular, Donnachie and Landshoff (DL)
have promoted \cite{DL} an appealing and very simple Regge parameterization  
of the total hadronic cross sections in which
\beq
\sigma_{\rm tot}\,=\,X\(\frac{s}{s_0}\)^{\Delta}\,+\,
Y\(\frac{s}{s_0}\)^{-\eta}.
\eeq
The two key ingredients of this approach are the Regge trajectories
\beq
\alpha_R(t)\,=\,\alpha_R(0)\,+\,\alpha_R^\prime \, t \, ,
\eeq
where $\alpha_R(0)\,=\,1-\eta$ and 
the Pomeron trajectory, which dominates at high energies,
\beq
\alpha_P(t)\,=\,\alpha_P(0)\,+\,\alpha^\prime_P \, t \, ,
\eeq
where $\alpha_P(0) = 1+\Delta$. The DL study establishes universal values 
$\Delta\,= 0.0808$ and $\eta = 0.4525$. This study is supplemented by
the analysis of Block, Kang and White \cite{KANG} who determine the slope 
$\alpha^\prime_P $
of the Pomeron trajectory to be $\alpha^\prime_P = 0.2\GeV^{-2}$.

The same approach is also applicable to the analysis of real
photoproduction and of the proton DIS structure function \cite{ALLM}.  
While  the energy dependence of the photoproduction total cross section
follows the DL pattern, it has been observed \cite{F2}
that $F_2(x,Q^2)$ behaves, for small enough x, like
$x^{-\lambda}$, where $\lambda$ is slowly growing with $Q^2$.
The growth of $\lambda$ is associated with the
behavior of the gluon structure function in the small $x$ limit
\beq\label{LAM}
\lambda\,=\,\frac{\partial ln\(xG(x,Q^2)\)}{\partial \ln(1/x)}\, .
\eeq

It has been recognized for quite some time that the transition from the
predominantly soft real photoproduction $(Q^2=0)$ to the predominantly
hard DIS processes, with high enough $Q^2$, is smooth and 
gradual \cite{ALLM,CKMT,GLMuni,DL2p}. 
This observation, regardless of its theoretical interpretation, is 
evident once we examine
the energy dependence of $F_2(x,Q^2)$ in the small x limit with $Q^2$
ranging from zero to a few $\GeV^2$. It is also well known that real
photoproduction of light vector mesons, $\rho$ and $\omega$, is dominated
by a soft Pomeron exchange \cite{DLph}\cite{GLMph}, whereas photoproduction
of heavy vectors, $J/\Psi$ and $\Upsilon$, is well reproduced by a
perturbative QCD (pQCD)  calculation 
 \cite{RYSKIN,BRODSKY,RRML,MRT},
 where ${M_V^2}/{4}$ replaces $Q^2$ as a measure of the process hardness.
Although  ${M_V^2}/{4}$ is a discrete variable, while  
 $Q^2$ is a continuous DIS variable, it is interesting to check if the
transition from soft to hard photoproduction of vector mesons follows the
behavior pattern observed in DIS.  To this end, the study of $\phi$
photoproduction is particularly instructive, as 
${M_{\phi}^2}/{4} = 0.26\GeV^2$,
while we know that the energy dependence of $F_2$, with small x and $Q^2$
as low as $0.2 - 0.3\GeV^2$, is steeper than the energy dependence of
$\sigma_{\rm tot}(\gamma p)$.  

Our investigation is susceptible to both
experimental and theoretical uncertainties. Experimentally, a systematic
study of the integrated $\phi$ photoproduction cross section and the
forward differential cross section slope 
\cite{phi2,phi3,phi4,phi5}
are not very reliable as
different experimental groups have utilized different, not always
mutually consistent, methods to extract and relate  these quantities. 
To overcome this
difficulty, we have analyzed the measured differential cross sections
rather than integrated quantities. Even so, the two higher energy data
points \cite{phi4,phi5} are averaged over wide energy bins. This,
combined with the overall poor quality of the reported data, may make a 
detailed analysis non conclusive at this stage. Theoretically, since we
wish to utilize the gluon structure function at low $Q^2$, the calculation
of the hard component requires some clarifications. Technically, a pQCD
calculation of 
$[{d\sigma(\gamma p \rightarrow \phi p)}/{dt}]_{t=0}$ 
depends on our knowledge of the
gluon structure function at $Q^2 = 0.26\GeV^2$. Such information requires
an extrapolation of a given parton distribution below its initial
evolution threshold $Q_0^2$. For this purpose we adopt a linear
extrapolation which was successfully utilized in previous
calculations \cite{GLMuni,GLMN}. As we shall see, there is a
significant difference between the MRST \cite{MRST} and GRV98 \cite{GRV98}
input gluon distributions. We have chosen to use the GRV98 distribution 
and shall explain our motivations for doing so.

The purpose of this letter is to examine these issues in some detail 
from different points of view. We present an analysis of the existing 
$\phi$ photoproduction forward differential cross section data 
which suggests an
energy dependence which is steeper than the typical energy dependence
associated with the soft Pomeron \cite{DL}.  We then present a pQCD
calculation from which we deduce that the hard component is responsible
for about a quarter of the $\phi$ photoproduction amplitude in the forward
direction at presently available energies. We then proceed to show
that a coherent sum of the calculated pQCD amplitude and a conventional
soft Pomeron exchange contribution provides an excellent reproduction of
the available data 
\cite{phi2,phi3,phi4,phi5}.

\section{Data analysis}

Our data analysis investigates whether the $\phi$ photoproduction 
cross section follows a power dependence on the c.m. energy
$W$, and whether this power is larger than the value determined from 
the energy dependence of the total cross section. 
Following DL \cite{DL} and Block et al. \cite{KANG}, we expect the
$\phi$ photoproduction cross section to behave like $W^{4\Delta}$ and the
forward slope to behave like $4\alpha_P^\prime \ln W$. The analysis of $\phi$
photoproduction data is seemingly easy, as this process proceeds 
exclusively through
Pomeron exchange, since the various Regge exchanges cancel each other. 
The problem is that the published analysis \cite{DL} \cite{phi5} depends on a
comparison between 
integrated cross section data taken by different groups who
have utilized different, and not always mutually consistent, procedures. 
In addition, because the $\phi$ forward slope is shrinking, the
interpretation of the integrated cross section behaving as a fixed power
of $W$ is somewhat ambiguous. 
In order to bypass these difficulties, we have limited ourselves to the 
analysis of the individual 
differential cross sections ${d\sigma}/{dt}$ as reported by the
experimental groups 
\cite{phi2,phi3,phi4,phi5}.
 We have used data with $W > 6 \GeV$ corresponding to $x < 0.025$.

Fig. 1 shows that $({d\sigma}/{dt})_0$ in the available energy
range is, indeed, well fitted by an effective power of W,  
\beq
\(\frac{d\sigma(\gamma p \rightarrow \phi p)}{dt}\)_{t=0}\,=\,A\, W^{4\lambda}.
\eeq
Our best fit, for 5 data points with $6.7 \leq W \leq 70\GeV$,
has an excellent ${\chi^2}/{n.d.f.} = 0.22$, corresponding 
to $A = 0.76 \pm 0.09  $ and $\lambda = 0.135 \pm  0.012$. 
For comparison we show also a 
fit where we fix the power to its DL value $4\Delta = 0.3232$ and obtain 
$A = 1.21 \pm 0.06 $ with 
${\chi^2}/{n.d.f.} = 0.92$.
This is a lesser quality fit, but it cannot be 
discredited. Our inability to determine the power
unambiguously results from the big error coupled to the ZEUS high energy
data point \cite{phi5}. Even so, it is clear that an improvement of the
HERA data
point at $<W> = 70\GeV$ and relevant data at higher energies will
enable
us to conclusively distinguish between a DL type interpretation and ours.

\begin{figure}
\centerline{\epsfig{file=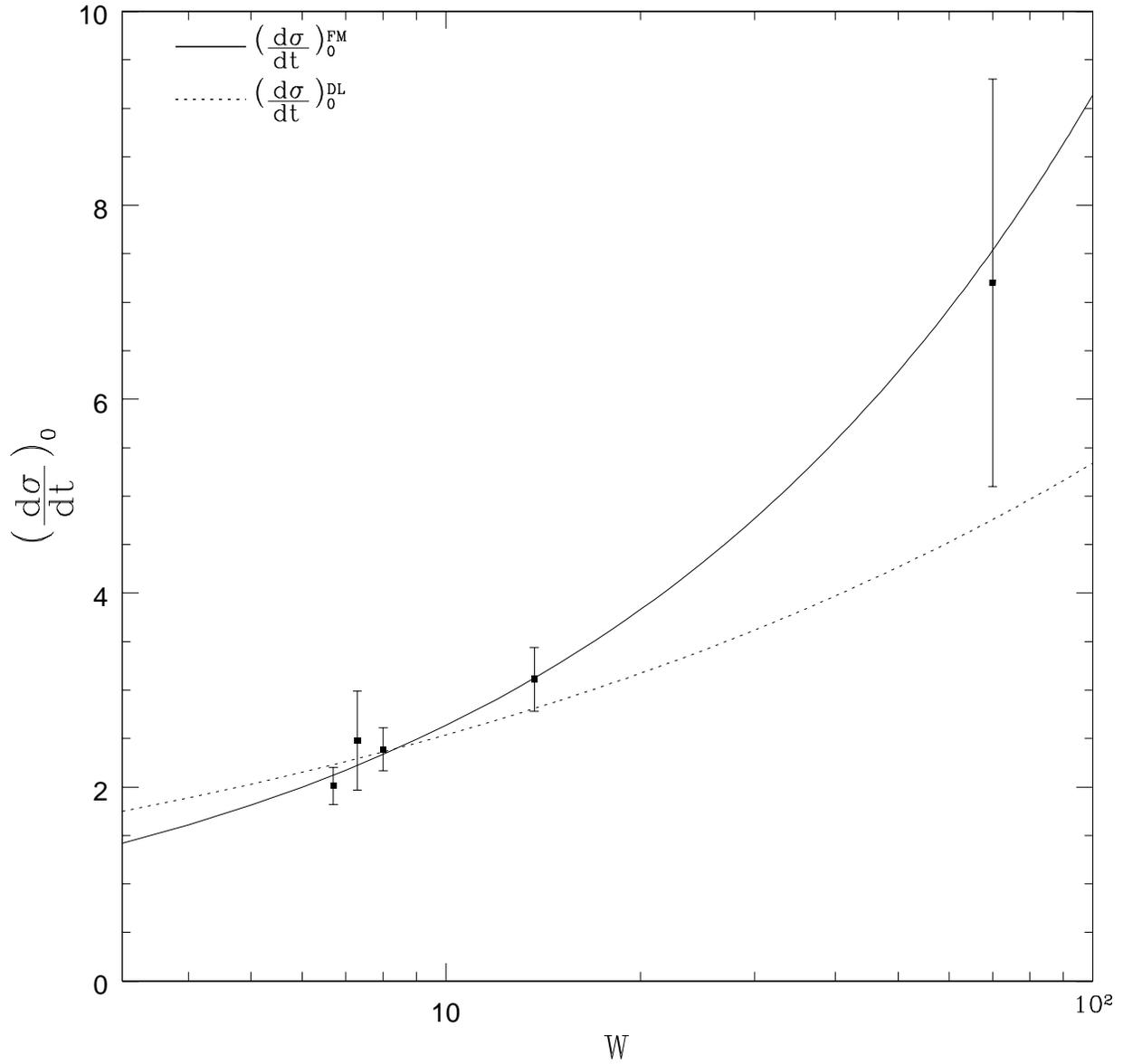,width=170mm}}
\caption{\it Best power fit compared with DL prediction
for $[{d\sigma(\gamma p \rightarrow \phi p)}/{dt}]_{t=0}$} .
\label{Fig.1}
\end{figure}

In order to further examine the suggestion that the dependence of $\phi$
photoproduction on $W$ is steeper than the behavior implied by a
soft Pomeron exchange, we have studied two (related) ratios
\beq 
R_1 = \frac{{(d\sigma}/{dt})_0}{\sigma_{\rm tot}^2(\phi p)} \,\, ,  
\eeq
and 
\beq
R_2 = \frac{({d\sigma}/{dt})_0}{\sigma_{P}^2(\gamma p)} \, \, ,
\eeq
where 
$\sigma_{P}(\gamma p)$ is the soft Pomeron exchange contribution to 
$\sigma_{\rm tot}(\gamma p)$. Both ratios are shown in Fig. 2. A careful
study of these ratios is of interest as both $\sigma_P^2(\gamma p)$ and 
$\sigma_{\rm tot}^2(\phi p)$ behave as $W^{4\Delta}$. Using the DL
parameterization \cite{DL} we have 
\beq
\sigma_P(\gamma p)\,=\,67.7 \, \( {\frac{W}{W_0}}\)^{0.1616}\mu b \,  
\eeq
and, with aid of the additive quark model relation between 
$\phi p$, $Kp$ and $\pi p$ cross sections, 
\beq
\sigma_{\rm tot}(\phi p)\,=\,10.01\,  \( {\frac{W}{W_0}} \)^{0.1616}\,+
1.51\, \({\frac{W}{W_0}}\)^{-0.4525} \, {\rm mb} \, ,
\eeq  
with $W_0 = 1\GeV$. 
A predominantly soft 
production mechanism for $\gamma p \rightarrow \phi p$ means that both
ratios presented in Fig. 2 are constants.
The best fits that we have obtained imply that $R_1$ behaves as 
$W^{0.278 \pm 0.086}$ and $R_2$ behaves as $W^{0.215 \pm 0.020}$. 
Constant ratios provide marginally acceptable fits to $R_1$ and $R_2$, but
this option cannot be definitely excluded.

\begin{figure}
\centerline{\epsfig{file=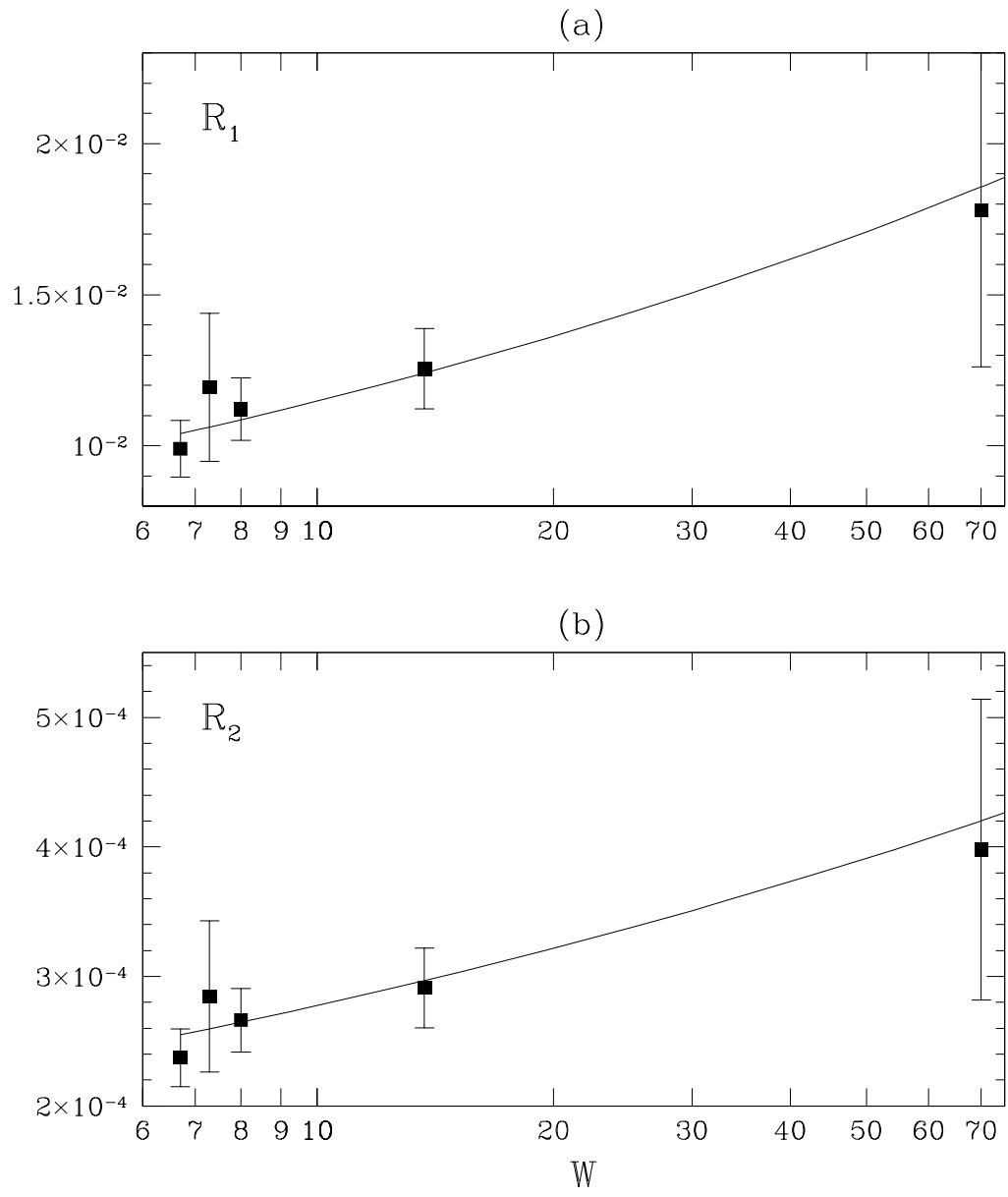,width=170mm}}
\caption{\it  $R_1$ and $R_2$ data and best fits.}
\label{Fig.2}
\end{figure}

To summarize: The available experimental data on
$({d\sigma(\gamma p \rightarrow \phi p)}/{dt})_0$ 
is consistently well described as having  
an energy dependence steeper than the one implied by a soft Pomeron
exchange. Nevertheless, the assumption of a pure soft production mechanism
cannot be unambiguously eliminated.

\section{A pQCD calculation}

Our pQCD calculation of the forward $\phi$ photoproduction follows
earlier 
pQCD calculations of the forward photoproduction cross section of heavy
vector mesons \cite{RYSKIN,BRODSKY,RRML,MRT}. 
These calculations are considerably simplified once we assume a
non-relativistic wave function for those vector meson states. This 
assumption, which is also valid for $\phi$, 
enables us to write a leading-order expression
\beq\label{LO}
\bigg[ \frac{d\sigma(\gamma p\rightarrow \phi p)}{dt}\bigg]_{t=0} = 
\frac{\alpha_S^2 \Gamma^{\phi}_{ee}}{3 \alpha_{EM} M_{\phi}^5}
16{\pi}^3
\bigg[ xG\(x,\frac{M_{\phi}^2}{4}\)\bigg]^2,
\eeq 
where $x = ({M_{\phi}}/{W})^2$ and $\Gamma^{\phi}_{ee}$ is the
partial
decay width of $\phi \rightarrow e^+e^-$. In the following we follow 
Refs. \cite{RRML}\cite{MRT} and, after calculating the cross
section resulting from the imaginary forward amplitude in leading order,
we correct for the real part of the amplitude and for higher orders.

The basic problem with our suggested calculation is that we depend on the
knowledge of the gluon structure function $xG(x,Q^2)$ at 
$Q^2 = M_{\phi}^2/4 = 0.26\GeV^2$. This $Q^2$ value 
is well below $Q_0^2$, the initial evolution
threshold, used in the updated parton 
distribution parameterizations \cite{MRST}\cite{GRV98} which took into 
account the behavior of $F_2$ and its logarithmic derivatives at small
$Q^2$. We recall the general property of the
gluon structure function 
which is linear in $Q^2$ in the limit of very small virtuality and
use, accordingly, a linear extrapolation
\beq
xG(x,Q^2)\,=\,\frac{Q^2}{Q_0^2}\, xG(x,Q_0^2) \,  \, \, , \, \, \, 
                                Q^2<Q_0^2 \, \, .
\eeq
This approximation has been successfully used in previous
theoretical DIS investigations \cite{GLMuni}\cite{GLMN} 
in which the knowledge of $xG$ in the small
$Q^2$ region was required. 

The results of our gluon structure function extrapolation for
MRST \cite{MRST} and GRV98 \cite{GRV98}, 
at the relevant $Q^2 = 0.26\GeV^2$
value, are shown in Fig. 3. Note that $Q_0^2 = 1.2 \GeV^2$ for MRST and 
$0.8\GeV^2$ for GRV98. Fig. 3 shows also the Pomeron term of
ALLM97 \cite{ALLM}. In the following we have used the GRV98 extrapolated
distribution. Our motivation is double folded:
\newline 
1) From a practical point of view, 
the MRST gluon structure function is considerably smaller than GRV98.
If we adopt the MRST distribution, we get a diminishing small hard
contribution for $\phi$ photoproduction in the $W$ range of interest. If
our data analysis is substantiated, we would then be left with the need 
for
some explanation for the observed energy dependence of the $\phi$ data. 
\newline
2) Theoretically, we note that MRST is very close to ALLM97 in the $x$
region of interest. Our interpretation is that the MRST input at small
$Q^2$ is predominantly soft, such as is ALLM97, and thus, even though 
perfectly legitimate, it is less suitable for our analysis.
\newline 

\begin{figure}
\centerline{\epsfig{file=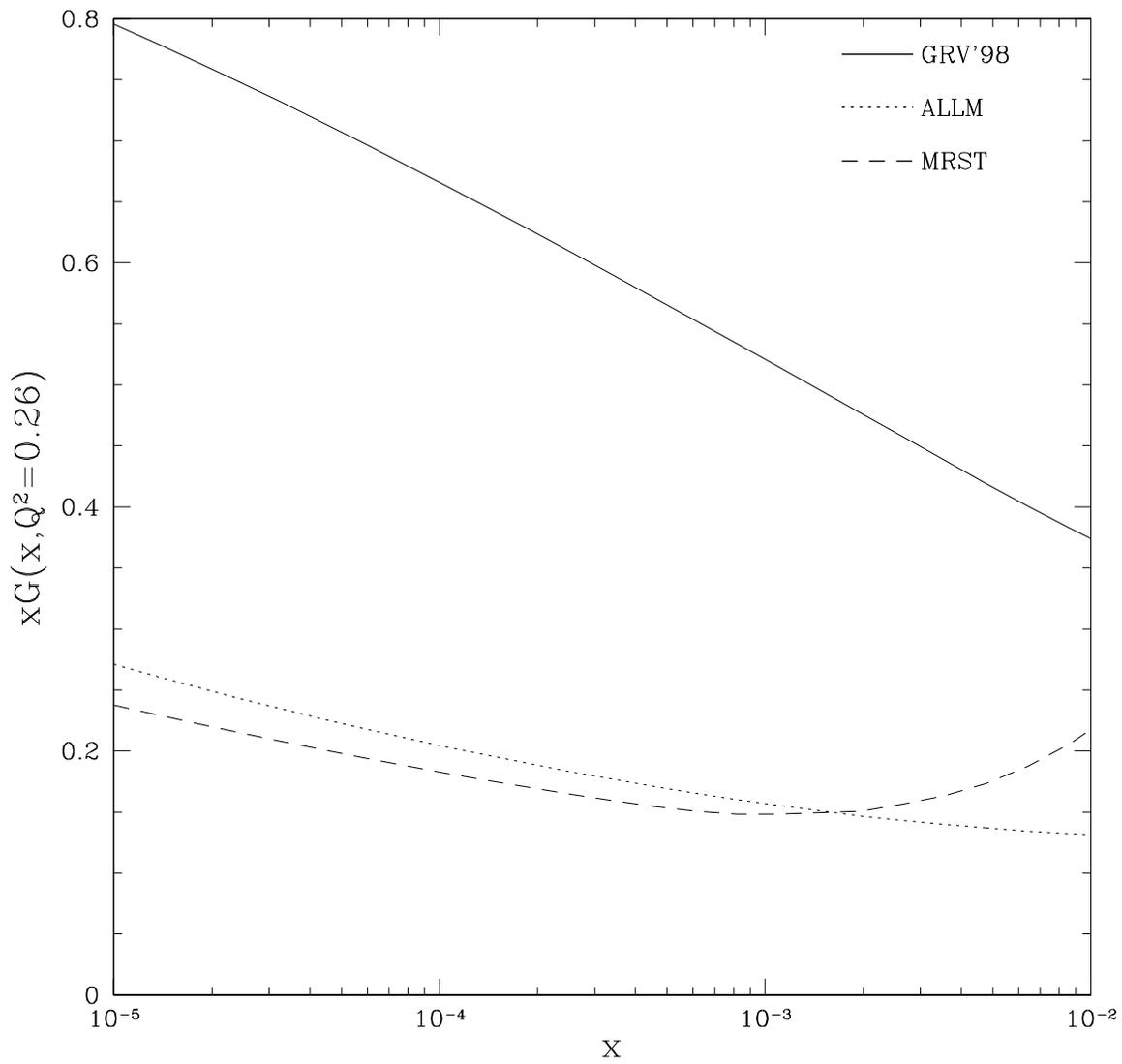,width=160mm}}
\caption{\it ALLM97 and the extrapolated MRST and GRV98 parameterizations
for the gluon structure function $xG(x,Q^2)$ at $Q^2\,=\,0.26\GeV^2$. }
\label{Fig.3}
\end{figure}

We follow Ref.\cite{RRML,MRT} and consider the following corrections
to the leading order cross section written in \eq{LO}:
\newline
1) \eq{LO} corresponds to the imaginary part of the forward amplitude and
should be corrected by $(1 + \rho^2)$, where 
$\rho = {\rm Re} A/{\rm Im} A$, and  
$A$ is the amplitude under consideration. We recall that 
$\rho \simeq {\pi \lambda}/{2}$, where, in our case, $\lambda$ is
given
by \eq{LAM}. We note that $\lambda = 0.145$ for $W = 70\GeV$ and
decreases
very slowly as the energy decreases. 
\newline
2) The next to leading order corrections are estimated by 
$[1 + 0.5 \alpha_S(M_{\phi}^2/4)]$.
\newline
3) Relativistic corrections and the effects of intermediate off diagonal
partons in $\phi$ photoproduction are rather small and have been
neglected.

With these effects put together, our overall correction factor is 
$C = 1.21$ at the high energy end of our 
data, decreasing slowly to $1.16$ in the lowest energy. The energy 
dependence of our calculated hard cross section is 
given in Table \ref{Table} and displayed in Fig. 4.

\section{A hybrid pQCD and soft Pomeron model}

Although the pQCD contributions calculated in the previous section 
provide a significant contribution to $\phi$ photoproduction, there
is no doubt that, in the energy range under consideration, the
leading production mechanism is a soft Pomeron exchange. Accordingly, we 
attempt to fit the data with a simple hybrid two- component model
with the following prescriptions:
\newline
1) The first component is a soft DL Pomeron with an 
$\alpha_P(t)$ intercept of 0.0808, namely 
\beq
\(\frac{d\sigma}{dt}\)^{\rm Soft}_{t=0}\,=\,A_S^2\(\frac{W}{W_0}\)^{0.3232}.
\eeq
\newline
2) A hard pQCD component $\(\frac{d\sigma}{dt}\)^{\rm Hard}_{t=0}$ , 
as calculated in the previous section.
\newline
3) A Coherent sum of the two component amplitudes.

We fit the 5 data points of $({d\sigma}/{dt})_0$ with one 
parameter, the normalization $A_S$ of the DL Pomeron. We obtain a best fit  
value $A_S =[ 0.83\pm0.02] (\mu b)^{0.5}$ with ${\chi^2}/{n.d.f.}= 0.54$. 
Our fitted cross sections (called FM) are presented in Table \ref{Table} and 
displayed in Fig. 4, both table and figure showing  also the pQCD and soft 
cross sections and our predictions for higher HERA energies. Our fit should 
be compared with the power fit, presented earlier,  which has a 
${\chi^2}/{n.d.f.} = 0.22$ and the
conventional DL fit which has ${\chi^2}/{n.d.f.} = 0.92$. Clearly, the 
presently available data is not sufficient to rule out any of these
options. Once again we note that this ambiguity results from the big error
associated to the $<W> = 70\GeV$ point. Improvement of  the quality of 
this point and additional HERA data will enable a more discriminative
analysis.

\begin{table}
\begin{center}
\begin{tabular}{||c|l|l|l|l||}
\hline\hline
\multicolumn{1}{||c|}{$W$} &
$\left(\frac{d\sigma}{dt}\right)_0^{\rm exp}$ &
\multicolumn{1}{c|}{$\left(\frac{d\sigma}{dt}\right)_0^{\rm FM} $} &
\multicolumn{1}{c|}{$\left(\frac{d\sigma}{dt}\right)_0^{\rm soft} $} &
\multicolumn{1}{c||}{$\left(\frac{d\sigma}{dt}\right)_0^{\rm pQCD} $}
\rule{0ex}{3ex}\rule[-2ex]{0ex}{2ex}\\
\hline\hline
6.7 & $2.01 \pm 0.19$ & 2.20  & 1.28  & 0.13 \\
7.3 & $2.48 \pm 0.51$ & 2.26  & 1.30  & 0.13 \\
8.0 & $2.39 \pm 0.22$ & 2.35  & 1.34  & 0.14 \\
13.7& $3.11 \pm 0.33$ & 2.91  & 1.60  & 0.20 \\
30  &                 & 3.96  & 2.06  & 0.31 \\
50  &                 & 4.79  & 2.43  & 0.40 \\
70  & $7.20 \pm 2.10 $ & 5.40  & 2.71  & 0.46 \\
100 &                 & 6.10  & 3.04  & 0.53 \\
150 &                 & 6.97  & 3.46  & 0.61 \\
200 &                 & 7.65  & 3.80  & 0.67 \\
250 &                 & 8.20  & 4.08  & 0.71 \\
\hline\hline
\end{tabular}
\caption{$\({d\sigma}/{dt}\)_{t=0}$ data and calculations.}
\label{Table}
\end{center}
\end{table}

\begin{figure}
\centerline{\epsfig{file=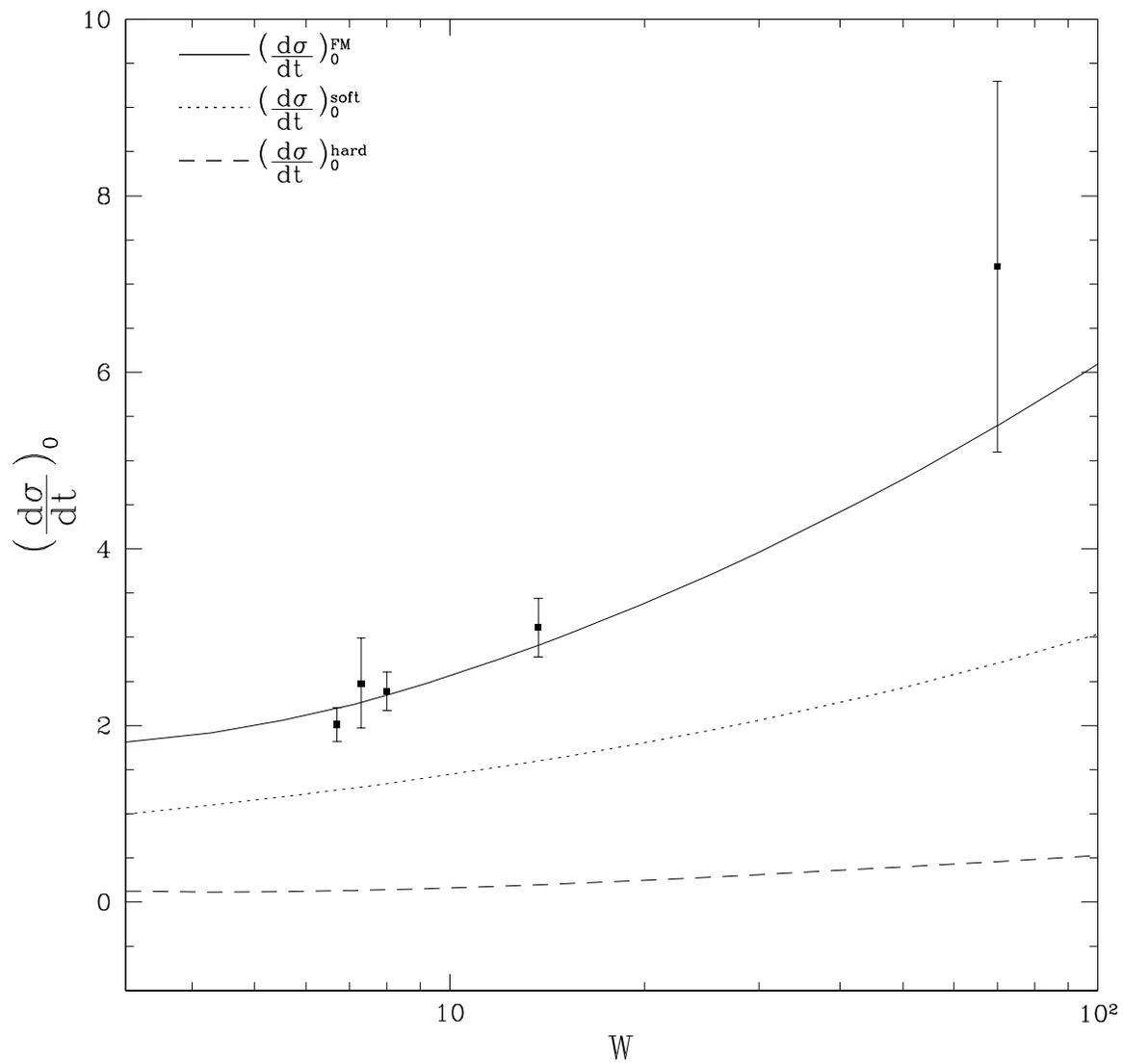,width=160mm}}
\caption{\it Data and our calculation for $({d\sigma}/{dt})_{t=0}$.
Also shown are our separate calculations of the soft and hard cross 
sections. }  
\label{Fig.4}
\end{figure}

The hybrid model that we have just suggested can be further examined by
considering the differential cross sections. Such data is
available \cite{phi4,phi5} at $<W> = 13.7$ and $70\GeV$.   
For the purpose of our calculations we need to know the  
t-dependence of both the pQCD and soft Pomeron amplitudes. To this end we
assume each of these dependences to be approximated by an exponential, 
\beq
F_i\,=\,F_i(0)e^{\frac{B_i}{2}t}\,\,\,\,\,\,i=S,H\,.
\eeq
This simple approximation is sufficient here, considering the quality of 
the available data on $\phi$ photoproduction. We then proceed as follows:
\newline
1) We take for the pQCD amplitude an energy independent exponential slope. 
We derive its value from the high energy differential
cross section on $J/\Psi$ photoproduction combined with the
observation \cite{LEVY} that this slope is energy independent,
corresponding to a flat hard Pomeron. In our analysis we have taken 
$B_H = 4.6\GeV^{-2}$, which corresponds to the H1 
measurement \cite{JH}. An equally good fit is obtained also with the ZEUS 
value \cite{JZ} of $B_H = 4.0\GeV^{-2}$. We have also treated $B_H$ as a
free parameter.
\newline 
2) For the soft Pomeron amplitude we assume a conventional Regge type
exponential slope depending on two fitted parameters,  
\beq
B_S\,=B_0^S\,+\,2\alpha_P^\prime \ln\(\frac{W}{W_0}\)^2 \, .
\eeq
The $B_0^S$ approximate value is known from the
phenomenology of $\phi$ photoproduction at low energies and 
$\alpha_P^\prime$ is approximately known from high energy
hadron-hadron phenomenology \cite{KANG}. As we show below, the fitted 
values
of these parameters are in excellent agreement with our expectations.

This hybrid model was fitted to reproduce 13 
$({d\sigma}/{dt})$ data points measured at $<W> = 13.7$ and $70\GeV$. 
We obtain $B_0^S = 4.20 \pm 0.90 \GeV^{-2}$ and 
$\alpha_P^\prime = 0.20 \pm 0.08$ with 
${\chi^2}/{n.d.f.} = 0.65$. Our results are in an
excellent agreement with soft hadron Regge analysis \cite{KANG}. Our soft
exponential slopes extrapolate well into the predominantly soft low
energy experimental data on $\phi$ phtoproduction \cite{phi1}. The data and
our fit are shown in Fig. 5. A slight improvement of the fit is obtained if
we take $B_H$ as a free parameter and obtain 
$B_H= 3.50 \pm 0.60\GeV^{-2}$, with the other parameters changing
insignificantly.

\begin{figure}
\centerline{\epsfig{file=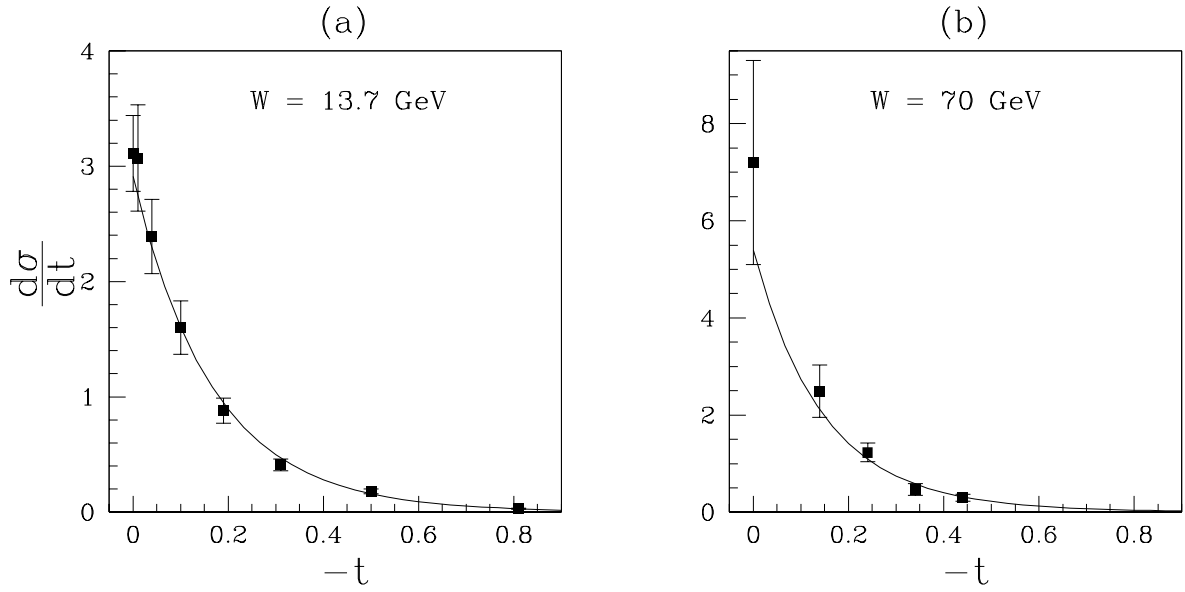,width=170mm}}
\caption{\it Data and our calculations for $({d\sigma}/{dt})$           
at $<W> = 13.7\,\, and\,\, 70 \GeV$.}
\label{Fig.5}
\end{figure}

Although our $({d\sigma}/{dt})$ fit corroborates our proposition
that high energy $\phi$ real photoproduction has a significant hard
component, we do not consider our success to be decisive. 
The $<W> = 70\GeV$ data, on its own, can be equally well described by
a single DL type soft component. Our two component model is appropriate to
describe the $<W> = 13.7\GeV$ data including $t > 0.4\GeV^2$, where a
single exponent is not sufficient. However, we caution against reaching 
too strong a conclusion from a single  measurement.
Clearly, and not only for the purpose of our analysis, additional
knowledge on higher $t$ behavior will help to clarify the picture.

\section{Discussion}

Following are the main conclusions of our study and some general remarks:
\newline
1) The data analysis of forward real $\phi$ photoproduction suggests the
existence of a significant hard pQCD contribution in the energy range of 
$6 \leq W \leq 70\GeV$. 
\newline
2) The above suggestion is corroborated by a pQCD calculation of 
$[{d\sigma(\gamma p \rightarrow \phi p)}/{dt}]_0$ using the GRV98
gluon distribution extrapolated to $Q^2 = 0.26\GeV^2$.
\newline 
3) A hybrid model, in which we coherently add the parameter free pQCD hard
component and a DL type soft component, provides an excellent overall
reproduction of the data. We note that the ratio of hard to soft
contribution in our model is compatible with the ratio obtained in a 
detailed study \cite{GLMuni} of DIS in comparable small $Q^2$ values.
\newline
4) Our hybrid model is significantly different from the two Pomeron model
suggested in Ref. \cite{DL2p}. In our model the $t = 0$ intercept of the
hard effective Pomeron is given by \eq{LAM} and as such it is a moving
pole. In the model of Ref. \cite{DL2p}, the hard Pomeron is a fixed pole
with a comparatively high $\lambda$.
\newline
5) Theoretically, the validity of our calculation rests on (i) the 
legitimacy of our $Q^2$ extrapolation of $xG(x,Q^2)$ below $Q_0^2$ ; and 
(ii) our choice of GRV98 for the input gluon structure function.
\newline
6) Our overall analysis strongly supports the existence of a significant
hard component contributing to $\gamma p \rightarrow \phi p$. However, a
decisive quantitative conclusion depends on improving and extending the 
HERA data on $\phi$ photoproduction.
\newline

{\Large \bf Acknowledgements:} 
We thank Eran Naftali for his help in the numerical computations.
UM wishes to thank UFRJ and FAPERJ (Brazil) for their support.
This research was supported by in part by the Israel Academy of Science
and Humanities.

\end{document}